\documentclass[pra,showpacs,superscriptaddress,twocolumn,10pt]{revtex4-1}
\usepackage{hyperref}
\usepackage[cm]{fullpage}
\usepackage{graphicx}
\usepackage[english]{babel}
\usepackage{amsmath}
\usepackage{amssymb}
\usepackage{cancel}
\usepackage{natbib}
\usepackage{comment}
\usepackage{color}
\usepackage[normalem]{ulem}

\newcommand{\ket}[1]{{\left\vert {#1} \right\rangle}}
\newcommand{\bra}[1]{{\left\langle {#1} \right\vert}}

\newcommand{\eq}{Eq.~}
\newcommand{\eqs}{Eqs.~}
\newcommand{\fig}{Fig.~}

\newcommand{\cf} {cf.~}
\newcommand{\ug} {\!=\!}

\newcommand{\piu} {\!+\!}
\newcommand{\meno} {\!-\!}

\newcommand{\eg} {e.g.~}

\newcommand{\rref} {Ref.~}
\newcommand{\rrefs} {Refs.~}

\makeatletter
	\renewcommand{\maketag@@@}[1]{\hbox{\m@th\normalsize\normalfont#1}}
\makeatother

\begin{document}

\author{Salvatore Lorenzo}
\affiliation{Quantum Technology Lab, Dipartimento di Fisica, Universit$\grave{a}$  degli Studi di Milano, 20133 Milano, Italy \& INFN, Sezione di Milano, I-20133 Milano, Italy,}
\author{Francesco Ciccarello}
\affiliation{NEST, Istituto Nanoscienze-CNR and Dipartimento di Fisica e Chimica, Universit$\grave{a}$  degli Studi di Palermo, via Archirafi 36, I-90123 Palermo, Italy}
\affiliation{Department of Physics, Duke University, P.O. Box 90305, Durham, North Carolina 27708-0305, USA}
\author{G. Massimo Palma}
\affiliation{NEST, Istituto Nanoscienze-CNR and Dipartimento di Fisica e Chimica, Universit$\grave{a}$  degli Studi di Palermo, via Archirafi 36, I-90123 Palermo, Italy}

\begin{abstract}
A collision model (CM) is a framework to describe open quantum dynamics. In its {\it memoryless} version, it models the reservoir $\mathcal R$ as consisting of a large collection of elementary ancillas: the dynamics of the open system $\mathcal{S}$ results from successive ``collisions" of $\mathcal{S}$ with the ancillas of $\mathcal R$. 
Here, we present a general formulation of memoryless {\it composite} CMs, where $\mathcal S$ is partitioned into the very open system under study $S$ coupled to one or more auxiliary systems $\{S_i\}$. Their composite dynamics occurs through internal $S$-$\{S_i\}$ collisions interspersed with external ones involving $\{S_i\}$ and the reservoir $\mathcal R$.
We show that important known instances of quantum {\it non-Markovian} dynamics of $S$ -- such as the emission of an atom into a reservoir featuring a Lorentzian, or multi-Lorentzian, spectral density or a qubit subject to random telegraph noise -- can be mapped on to such {\it memoryless} composite CMs. 
\end{abstract}


\title{Composite quantum collision models}

\date{\today}

\maketitle

\section{Introduction}

A longstanding problem in the field open quantum system dynamics is the derivation of an effective description of the reduced dynamics of a system $\mathcal{S}$ in contact with the surrounding environment, i.e., of a master equation having the reduced density operator of $\mathcal{S}$ as the only unknown \cite{open,open1,open2}. This is in general a highly non-trivial task for  quantum non-Markovian dynamics. Note that even the very meaning non-Markovianity and its correct measure are currently the focus of intense investigations \cite{NM}. Sometimes the approximations made to describe non-Markovian dynamics can lead to master equations (MEs) which do not preserve trace and complete positivity.

Quantum  collision models, first introduced in \cite{Rau} and more recently studied in \cite{CMs,buzek} have proved to be a promising tool to analyse quantum non-Markovian dynamics \cite{rybar,Palma2012,ciccarello,ruari,diosi,santos} as well as of quantum thermodynamical systems (see \eg \rrefs \cite{landauer, flux, thermo}).
In its standard, {\it memoryless}, version, a collision model describes the reservoir as a large collection of elementary constituents or ``ancillas" and the joint dynamics as a sequence of pairwise system-ancilla unitary ``collisions". 
The resulting reduced {non-unitary} dynamics of $\mathcal{S}$, in the continuous-time limit, can be shown \cite{buzek} to be described by a Gorini-Kossakowski-Sudarshan-Lindblad (GKSL) master equation  \cite{open,open1,open2}. A memoryless collision model thus entails a fully Markovian evolution for the open system as long as the ancillas are initially in a {\it product} state, they do not mutually interact and the system collides only once with each of the ancillas. To account for non-Markovian processes in collision  models, one has to somehow relax such assumptions, \eg by allowing the initial reservoir state to be correlated \cite{rybar,santos} or enabling inter-ancillary interactions between next system-ancilla interactions \cite{ciccarello, ruari}. 

Collision model-based approaches are promising for the study of non-Markovian dynamics for at least two reasons: $(i)$ they allow for the possibility to decompose a complicated open dynamics in terms of discrete elementary processes (each usually involving a pair of low-dimensional systems) and $(ii)$ they make possible the derivation of well-behaved non-Markovian master equations \cite{ciccarello,Palma2012}. Remarkably, $(i)$ in particular can suggest schemes to perform experimental simulations of non-Markovian dynamics \cite{sciarrino}  or provide valuable theoretical tools in the analysis of time-delayed quantum feedback \cite{grimsmo}.

An open issue in quantum collision models is their descriptive power. While being an advantageous tool in many respects, a collision model is itself rather abstract. One thus naturally wonders whether (and how), given a non-Markovian open dynamics, this can be reproduced through a suitably built collision model.  In the case of a qubit, Rybar {\it et al.} showed that any non-unitary dynamics can be described through a collision model provided that the initial reservoir state is chosen accordingly. Typically, however, this requires the preparation of a multi-partite correlated state of all the reservoir ancillas, which may be an experimentally demanding task. 
Concerning collision models with initial uncorrelated reservoir states, instead, very few instances of non-Markovian dynamics were so far demonstrated to be reproducible through a collision model. 

A recurrent situation in which a non-Markovian dynamics emerges (e.g.~in quantum optics or condensed matter scenarios) is when the interaction between a small quantum system $S$ and a large reservoir $\mathcal R$ is not direct but bridged by an ``auxiliary" quantum system $S_1$ \cite{JC, Anderson}. The prototypical tool for describing such dynamics is a GKSL master equation for the joint state of $S$ and $S_1$, where the Hamiltonian term of the Lindblabian superoperator features in particular a direct coherent $S{-}S_1$ coupling while the non-Hamiltonian one depends on a set of jump operators defined in the Hilbert space of $S_1$ only. While the resulting joint dynamics of $S{-}S_1$ is evidently Markovian, the one of $S$ is in general non-Markovian.
The question is now raised whether a collision model effectively describing the evolution of $S$ in the continuous-time limit can be defined for such an important class of non-Markovian dynamics. 

One is thereby intuitively led to consider a memoryless collision model (non-interacting and initially uncorrelated ancillas) where however the system $\mathcal S$ undergoing repeated collisions with the ancillas is now {\it multipartite}, being composed by $S$ (the very open system under study) and an auxiliary system $S_1$. Natural requirements would be to let $S$ be uncoupled from the reservoir, but allow for a direct $S$-$S_1$ coherent interaction to occur between collisions. The main aim of the present paper is to formulate in a rigorous way a theoretical framework showing that it is indeed possible to define a family of quantum collision models -- which we call {\it composite collision models} -- that are precisely based on this intuitive idea and reproduce the class of non-Markovian dynamics described above. The discrete dynamics of such models can be thought as consisting of  ``internal" collisions -- enabling a crosstalk between $S$ and $S_1$ -- interspersed with collisions between the auxiliary system $S_1$ only and the reservoir ancillas.
The effectiveness of this framework is illustrated in the case of some specific instances of composite collision models, showing in particular that \eg the known non-Markovian decay of an atom in a lossy cavity or the dynamics of a qubit subject to random telegraph noise can emerge through a collision model-based formulation. The collision models we introduce are naturally extended, as we show, to the case of a manifold of auxiliary systems $\{ S_1,S_2,...\}$.

The use of a  bipartite collision model to describe a damped Jaynes-Cummings-model dynamics was introduced in \rref\cite{landauer} and then investigated in more detail in \rref\cite{luoma}. Here, this result emerges as a specific instance of our composite collision model framework. In particular, we present a thorough discussion of the conditions to match in order for such effective description to hold in the continuous-time limit.

On a rather general ground, any open system dynamics of a system $S$ arises as the partial trace over the environmental degrees of freedom of the joint unitary dynamics entailed by the system-reservoir total Hamiltonian model \cite{open,open1,open2}. Such environmental model has on the one hand a clear physical meaning while, on the other hand, allows for a joint dynamics where a large number of degrees of freedom are involved. In contrast, a collision model dynamics takes place through a succession of elementary interactions -- each involving only a small reservoir subunit -- but its connection to a realistic physical scenario is less
 straightforward. In the light of this, given a microscopic environmental model, it would be highly desirable to devise a general method to associate a collision model yielding the same open system dynamics in the continuous-time limit. Here, we take a first step towards this challenging goal by showing that such mapping is possible for some specific environmental models. This can be the case for a qubit that is coupled in a purely dissipative or dispersive fashion to a bosonic bath when the spectral density has a Lorentzian or multi-Lorentzian shape, as we show.

The outline of this paper is the following.  In Section \ref{review}, we review the standard quantum collision model leading to a GKSL master equation in its continuous-time limit. In Section \ref{MCM-internal}, we show how and under what conditions the collision model of Section \ref{review} can be extended to include an internal system dynamics described by a corresponding free Hamiltonian. The theoretical framework so formulated is then used in Section \ref{MCM-multipartite} as the basis to define a composite quantum collision model in the bipartite case. In Section \ref{lossy},  we illustrate a prominent instance of such models which in the continuous-time limit effectively reproduces the open dynamics of an atom decaying in a lossy cavity (damped Jaynes Cummings model).  In Section \ref{dephasing}, we study another instance of composite bipartite collision model based on a dispersive $S$-$S_1$ coupling, either with respect to $S_1$ or $S$. Correspondingly, the resulting collision model can describe either a qubit subjected to random-telegraph-noise or a qubit undergoing a purely dephasing dynamics. In Section \ref{MCM-multipartite-multi}, we show how to extend the composite bipartite collision model of Section \ref{MCM-multipartite} to the multipartite case. An instance, based on a tripartite collision model, is then presented in Section \ref{multi} and shown to be able to reproduce the dynamics of an atom dissipatively coupled to a reservoir featuring a SD that is the sum of two Lorentzian distributions.
Finally, in Section \ref{Conclusions} we draw our conclusions.

\section{ Memoryless collisional model and the markovian master equation}\label{review}

 In this section, we will briefly review how the standard Markovian GKSL master equation (ME) is naturally derived by a collisional memoryless model of open dynamics. In such a model a quantum reservoir $\cal{R}$ consists of a large ensemble of identical non-interacting ancillas $\{R_n\}$ all in the same initial state. The system  $\mathcal{S}$ interacts with the environment via a sequence of ``collisions" - i.e., short interactions -  with each of the ancillas. 
The initial joint state of  $\mathcal{S}$-$\mathcal{R}$ is assumed to be the product state
\begin{equation} 
\sigma_{0}\ug\rho_{0}\otimes\left(\eta\otimes\eta\otimes...\right)\,,\label{sigma0}
\end{equation}
where $\rho_0$ is the initial state of $\mathcal{S}$ while $\eta$ is the common initial state of all the ancillas. Both $\rho_0$ and $\eta$ can in general be mixed. 
The state $\eta$ can always be expressed in diagonal form in terms of its eigenstates $\{|m\rangle\}$ and associated probabilities $\{p_m\}$ as
  \begin{equation}
 \eta=\sum_m p_m\,|m\rangle\langle m|\,,\label{eta}
 \end{equation}
where $\{|m\rangle\}$ form an orthonormal basis of the ancilla Hilbert space. In the memoryless version of the model the reservoir is assumed to be so large that the system never collides twice with the same ancilla, therefore the open dynamics of $\mathcal S$ takes place through pairwise short interactions between $\mathcal S$ and each reservoir ancilla: $\mathcal S$-$R_1$, $\mathcal S$-$R_2$, $\mathcal S$-$R_3$,...$\,$ in such a way that at each step $\mathcal S$ collides with a ``fresh" ancilla that is still in state $\eta$. A schematic sketch of the model dynamics is given in \fig1(a).
 
 It is assumed that all the collisions have the same duration $\tau$, each being described by the unitary evolution operator $\hat U_{\mathcal{S}n }$ given by (we set $\hbar\ug1$ throughout)
 \begin{equation}
 \hat {U}_{\mathcal{S}n }=e^{-i\hat H_{\mathcal {S}n}\tau}\,,\label{Usn1}
 \end{equation}

with
 \begin{equation}
 \hat H_{\mathcal{S}n }=g \,\hat h_{\mathcal{S}n}\label{Hsn}
 \end{equation}
where $g$ is a coupling rate and $\hat h_{\mathcal {S}n}$ is a dimensionless Hermitian interaction hamiltonian acting on the joint $\mathcal S$-$R_n$ Hilbert space.
 
Let $\rho_{n}$ be the state of $\mathcal S$ at the (generic) $n$th step, i. e., just after the collision with the $n$th ancilla:
\begin{eqnarray}
\rho_{n}{=}
{\rm Tr}_{\mathcal R}\!\left\{\hat {U}_{\mathcal {S}n }\rho_{n-1}\,\eta\,\hat {U}_{\mathcal {S} n }^\dag\right\}\equiv{\rm Tr}_{R_n}\!\!\left\{\hat {U}_{\mathcal {S}n }\rho_{n-1}\,\eta\,\hat { U}_{\mathcal {S}n }^\dag\right\}\label{rhon}\,\,\,\,\,
\end{eqnarray}

\subsection*{Continuous-time limit}

As we assumed each collision to last for a short time $\tau$, we approximate $\hat {U}_{\mathcal{S}n}$ [\cf\eq(\ref{Usn1})] {up to the second order in $\tau$} as
 \begin{equation}
 \hat {U}_{\mathcal{S}n }\simeq \openone_{\mathcal{S}n}-i {\hat H_{\mathcal{S}n}}\tau-\frac{\hat{H}_{\mathcal{S}n }^2}{2}\tau^2\,.
 \end{equation} 
When this is substituted into \eq \eqref{rhon} the variation of $\rho_n$ due to a single collision, to second order in $\tau$, is
\begin{eqnarray}
{\Delta \rho_n}&\ug&\,{\rm Tr}_{R_n}\!\left\{-i\,[\hat{h}_{\mathcal {S}n},\rho_n\eta]\right\}\!g\tau\nonumber\\
&&+\, {\rm Tr}_{R_n}\!\!\left\{\hat{h}_{\mathcal{S}n}(\rho_n\eta) \,\hat{h}_{\mathcal{S}n}\!
\meno\frac{1}{2}\left[\hat{h}_{\mathcal{S}n}^2,\rho_n\eta\right]_+\right\}\!(g\tau)^2\label{drho}\,\,\,\,\,\,\,
\end{eqnarray}
with $\Delta\rho_n\ug{\rho_{n}\meno\rho_{n-1}}$, $[\hat C,\hat D]\ug\hat C\hat D\meno \hat D\hat C$ and $[\hat C,\hat D]_+\ug\hat C\hat D\piu \hat D\hat C$.
In line with standard procedures in open quantum system theory \cite{open,open1,open2}, it is also assumed \cite{Palma2012} that
\begin{equation}
{\rm Tr}_{R_n}\{ \hat h_{\mathcal {S}n}\eta\}=0\,.\label{average}
 \end{equation}
This assumption can be made with no loss of generality since, when the average (\ref{average}) is non-zero, it amounts just to a renormalization of the $\cal S$ Hamiltonian and can thereby be incorporated in the free-system Hamiltonian  \cite{Palma2012, cohen}.

 Let now $t_n\ug n\tau$ (with $n\ug0,1,...$) be the discrete time variable up to  the $n$th step. As one can equivalently regard the collision model as the interaction of $\mathcal S$ with only {\it one} ancilla, whose state is refreshed to $\eta$ at times $t_n$, the collision time here plays the role of the usual environment self-correlation time in standard microscopic derivations of the GKSL master equation \cite{open}. This time is, strictly speaking, finite. To pass from the discrete dynamics to the continuous-time one we must therefore realise that what we have in mind is a sort of coarse graining over a finite time. From a formal viewpoint, we carry out this by taking the limit $n\!\gg\!1$ and $\tau\!\simeq\!0$ in such a way that $t_n\!\rightarrow\!t$ with $t$ being now a continuous-time variable. Accordingly, $\Delta\rho_n/\tau\!\rightarrow\!d\rho/dt$. At the same time we assume that 
 the product $\gamma\ug g^2\tau$\label{gammaMCM} remains finite. Note that in microscopic derivations $\gamma$ is proportional to the self-correlation time. 
 \begin{figure}
\includegraphics[width=\columnwidth]{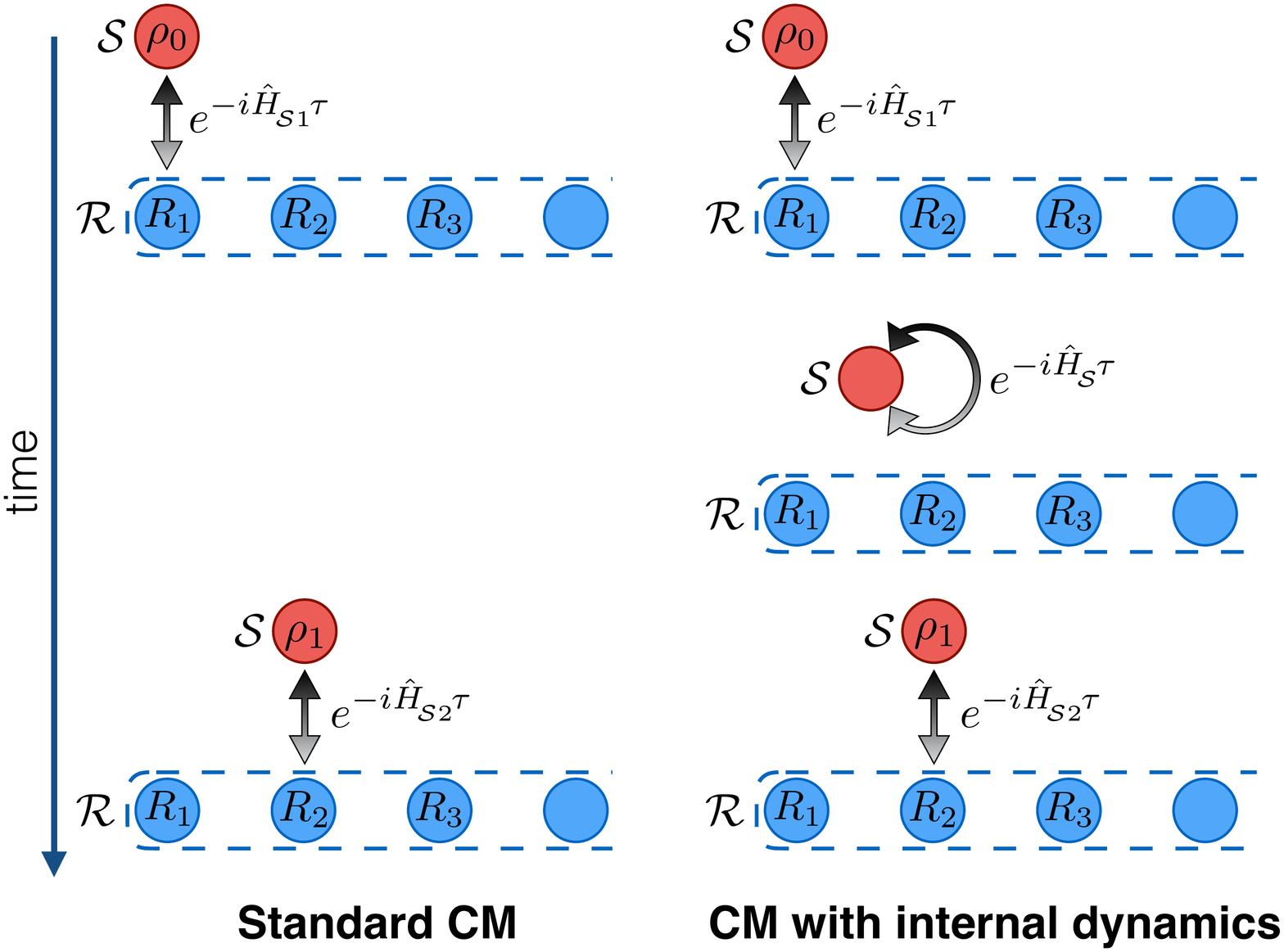}
\caption{(a) Standard memoryless collision model: the system $\mathcal S$ undergoes successive collisions with the reservoir ancillas $\{R_n\}$, each corresponding to an interaction Hamiltonian $\hat H_{\mathcal Sn}$. (b) Collision model with internal dynamics: unlike (a), an intra-system collision (corresponding to a free system Hamiltonian $\hat H_{\mathcal S}$) takes place between two next system-ancilla collisions. In either case of collision model, only the first steps are sketched, the next ones being obtained by simple iteration. \label{fig1}}
\end{figure}
In the continuous-time limit just described, thereby, the finite-difference equation (\ref{drho}) takes the form of a continuous-time master equation
\begin{equation}
\frac{{\rm d}\rho}{{\rm d}t}=\mathcal L(\rho) \label{ME1}
\end{equation}
with the superoperator $\mathcal L$ given by
\begin{equation}
\mathcal L(\rho)=\gamma\sum_{\mu\nu} \left(\hat{A}_{\mu\nu}\rho\hat{A}_{\mu\nu}^\dag\meno\frac{1}{2}[\hat A_{\mu\nu}^\dag\hat A_{\mu\nu},\rho]_+\right). \label{lindbladian}
\end{equation}
Here, $\{\hat{A}_{\mu\nu}\}$ are jump operators in the $\mathcal S$ Hilbert space defined by
\begin{equation}
\hat{A}_{\mu\nu}=\sqrt{p_\nu}\,\,
\!\langle \mu|\hat{h}_{\mathcal {S}n}|\nu\rangle\!
\label{Amunu} 
\end{equation}
where [\cf\eq(\ref{eta})] $|\mu\rangle$ and $|\nu\rangle$ are two orthonormal eigenstates of $\eta$, i.e., elements of the basis $\{|m\rangle\}$, and $p_\nu$ the $\nu$th eigenvalue of $\eta$ [owing to the collision model translational invariance, jump operators \eqref{Amunu} and thus $\mathcal L$ are independent of ancilla $R_n$].

\section{Memoryless collision model with internal dynamics}\label{MCM-internal}

The standard collision model of the previous section can be modified  to allow for an {\it internal dynamics} of $\mathcal S$ to take place as well. Specifically, we assume that, between two consecutive system-ancilla collisions, $\mathcal S$ undergoes a unitary dynamics governed by a free Hamiltonian $\hat H_{\mathcal S}$ as shown in \fig1(b). We will refer to this process as an ``intra-system collision".
A ``step" is now defined so to incorporate one intra-system collision, lasting a time $\tau_s$  followed by a system-ancilla one, lasting a time $\tau_n$.
The system evolution after the $n$th step is again described by \eq(\ref{rhon}), but  $\hat{{U}}_{\mathcal Sn}$ is now given by

 \begin{equation}
 \hat {U}_{\mathcal{S}n }=e^{-i \hat{H}_{\mathcal{S}n }\tau_n}\,e^{-i \hat{H}_{\mathcal S}\tau_s}\,\label{Usn2}\,,
 \end{equation} 
 where the $S$-$R_n$ interaction Hamiltonian $\hat H_{\mathcal {S}n}$ is the same as \eq(\ref{Hsn}) while 
 \begin{equation}
 \hat H_{\mathcal{S} }=J \,\hat h_{\mathcal S}\label{Hs}
 \end{equation}
 is the free Hamiltonian of $\mathcal S$ with characteristic frequency $J$ and where $\hat h_{S}$ is a dimensionless operator defined in the $\mathcal S$ Hilbert space. 
 In the following we want to reproduce a coherent dynamics, generated by $\hat h_{S}$, together with an incoherent dynamics, due to the system-ancilla collisions. To be consistent
 with this assumption, while we coarse grain on the incoherent dynamics, which means that $\tau_n$  will be assumed to be small but finite, we will assume $\tau_s \ll \tau_n$ 
Consistently, $\hat {{U}}_{\mathcal {S}n}$ in \eq(\ref{Usn2}) is approximated as
 \begin{equation}
 \hat {U}_{\mathcal{S}n }\simeq \openone_{\mathcal{S}n }-i (\hat{H}_{\mathcal S}\tau_s+{\hat H_{\mathcal {S}n}\tau_n})-\frac{\hat{H}_{\mathcal{S}n }^2}{2}\tau_n^2\,.\label{Usn3}
 \end{equation} 
 \eq(\ref{Usn3}) can be obtained by approximating in \eq(\ref{Usn2}) $e^{-i \hat{H}_{\mathcal{S}}\tau}$ up to {\it first} in  $\tau_s$
and  $e^{-i \hat{H}_{\mathcal{S}n }\tau}$  to {\it second} order order in $\tau_n$, 
respectively, and then neglecting terms $\sim\!\tau_s\tau_n^2$ as well as terms $ \tau_s\tau_n$. Note that, given the approximations made, in particular neglecting terms $\sim\!g J$, the two unitaries in \eq(\ref{Usn2}) commute: it is therefore irrelevant whether the system-ancilla collision occurs before or after the intra-system one. 
Thereby, one can equivalently regard the two elementary collisions as if they occurred simultaneously and assume
[\cf \eq(\ref{Usn2})], $\hat {U}_{\mathcal{S}n }\!\simeq\!e^{-i (\hat{H}_{\mathcal{S}}\piu \hat{H}_{\mathcal{S}n})\tau}$. 
In particular, this makes legitimate to set (see a few lines below) $\dot \rho\!\simeq\!\Delta\rho_n/\tau$ even if the time step consists of two subsequent collisions  duration $\tau_n$ and $\tau_s$.
\begin{figure}
\includegraphics[width=\columnwidth]{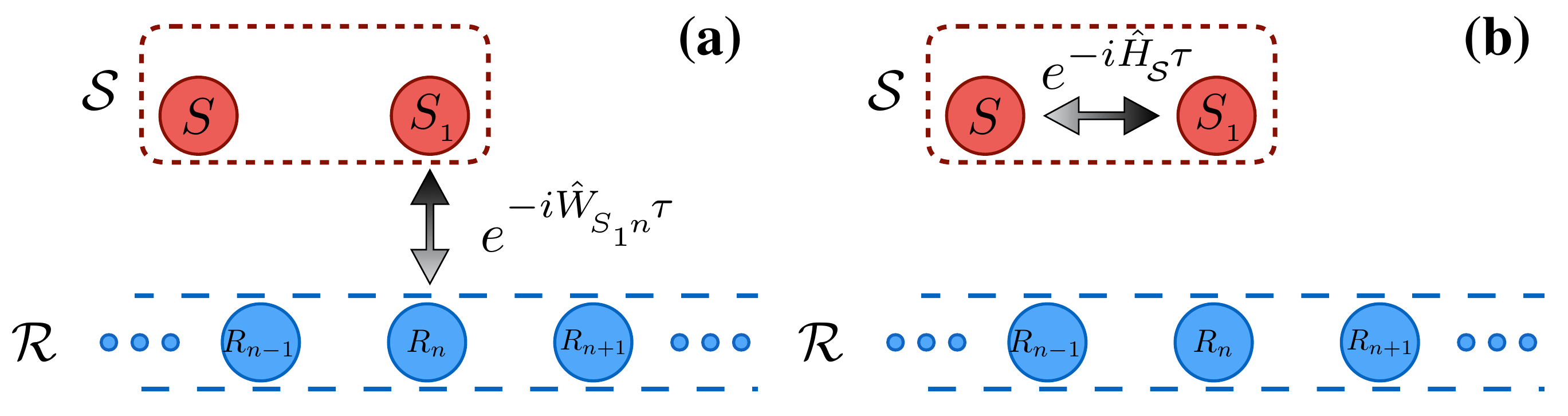}
\caption{System-ancilla collision (a) and intra-system collision (b) in a composite bipartite collision model. System $\mathcal S$ comprises the very open system $S$ under study and an auxiliary system $S_1$. In (a), note that only $S_1$ is involved in the collision with a reservoir ancilla. In (b), Hamiltonian $\hat H_{\mathcal S}$ in particular accounts for a direct $S$-$S_1$ interaction.  Apart from these specifications, the general discrete dynamics takes place analogously to \fig 1(b). \label{fig2}}
\end{figure}

Proceeding now in analogy with the previous section, we get the identity
\begin{eqnarray}
{\Delta \rho_n}&\ug&-i\,[\hat{H}_{\mathcal S},\rho_n]\tau\piu {\rm Tr}_{R_n}\!\!\left\{\hat{H}_{\mathcal {S}n}\rho_n\eta \hat{H}_{\mathcal {S}n}\right\}\tau^2\!\nonumber\\
&\meno&\frac{1}{2}{\rm Tr}_{R_n}\!\left\{[\hat{H}_{\mathcal {S}n}^2,\rho_n\eta]_+\right\}\tau^2\label{drho2}\,.
\end{eqnarray}
Correspondingly, in the continuous-time limit we end up the master equation
\begin{equation}
\frac{d\rho}{dt}=-i\,[\hat{\mathcal H}_{\mathcal S}, \rho]\mathcal+ \mathcal L(\rho) \label{ME}\,,
\end{equation}
where the superoperator $\mathcal L$ has the same form as in \eq(\ref{lindbladian}) with the associated jump operators given by \eq(\ref{Amunu}). 

We point out that, as in the previous model, $\mathcal S$ undergoes a Lindbladian (hence Markovian) dynamics and that the internal dynamics of $\mathcal S$ only appears in the Hamiltonian term of the right-hand side of \eq(\ref{ME}). This is a consequence of the fact that are treating the system-reservoir dynamics in a coarse-grained fashion, while the system's internal dynamics is taken into account in full detail.

Note that the collision model with no internal dynamics of Section \ref{review} is effective even in the presence of a system free Hamiltonian provided that $[\hat H_{\mathcal S},\hat H_{\mathcal S n}]\ug0$. In such a case,
it indeed corresponds to the interaction picture. If $[\hat H_{\mathcal S},\hat H_{\mathcal S n}]\!\neq\!0$, though, this is no longer true since the system-ancilla interaction Hamiltonian in Section \ref{review} is assumed to be time-independent. To avoid time dependancies regardless of such commutation relationship, the $\mathcal S$ internal dynamics thus must be explicitly involved in the collisional dynamics, as shown above.

\section{Composite collision model} \label{MCM-multipartite}

We are now ready the composite quantum collision model that is central to our study. This is in fact a specific instance of the collision model with internal dynamics analysed in the previous section, where $\mathcal S$ is a {\it bipartite} system (in Section \ref{MCM-multipartite-multi} we will discuss the extension to the multipartite case). Specifically, $\mathcal S$ comprises subsystems $S$ and $S_1$ (see \fig2) with $S$ embodying the very open system under study, while $S_1$ plays the role of an auxiliary system (note that in the collision model with internal dynamics of the previous section $\mathcal S\!\equiv\!S$). By definition, the free hamiltonian of $\mathcal S$ reads
\begin{equation}
\hat {H}_\mathcal S=\hat {H}_{S_1}+\hat V_{S\!S_1}\,,\label{HSC}
\end{equation}
where $\hat {H}_{S_1}$ is the free Hamiltonian of $S_1$ (the one of $S$ is assumed to be zero) and $\hat V_{S\!S_1}$ is the interaction Hamiltonian of $S$ and $S_1$.
As for the $\mathcal S$-$R_n$ interaction [\cf\eq\eqref{Hsn}], this takes the form 
\begin{equation}
\hat {H}_{\mathcal{S}n }=\hat {W}_{S_1 n}\ug g \,\hat {w}_{S_1 n} \label{HSn3}
\end{equation}
with $\hat {w}_{S_1 n}$ a dimensionless operator acting on the Hilbert space of subsystem $S_1$ and ancilla $R_{n}$ ($g$ is the associated coupling strength). System $S$ is thus not subject to any direct interaction with $R_n$. A sketch of the collision model dynamics is given in \fig2.

The master equation in the continuous-time limit thus reads
\begin{eqnarray}
\frac{d\rho}{dt}=-i\,[\hat{\mathcal H}_{\mathcal S}, \rho]\mathcal+\mathcal L_{S_1}(\rho)\label{MEC}
\end{eqnarray}
with $\mathcal L_{S_1}(\rho)$ having a form analogous to \eq(\ref{lindbladian}) with
\begin{eqnarray}
\hat{A}_{\mu\nu}^{(1)}&\!=\!&\sqrt{p_\nu}\,\,_{R_{n}}\!\langle \mu|\hat w_{S_1\!R_{1}}|\nu\rangle_{R_{n}},\label{Amunui}\\
\gamma&\!=\!&{g^2\tau}\label{gammai}\,.
\end{eqnarray}
The jump operators $\{\hat{A}_{\mu\nu}^{(1)}\}$ act in the Hilbert space of $S_1$.

In the next two sections, we discuss two important instances of bipartite composite collision model, showing their connection with known relevant classes of open quantum system dynamics.

\section{Atom in a lossy cavity}\label{lossy}

Based on the definitions in Section \ref{MCM-multipartite}, consider now the case where $S$ and $S_1$ are, respectively, a qubit and a bosonic mode. Let $\{\hat\sigma_\pm$, $\hat\sigma_z\}$ be the usual Pauli spin operators associated with $S$, while $\hat \alpha$ ($\hat \alpha^\dag$) is the annihilation (creation) bosonic operator for the auxiliary system $S_1$. The $n$th reservoir ancilla $R_n$ is modelled as as a bosonic mode with associated annihilation (creation) operator $\hat a_n$ ($\hat a_n^\dag$).
By definition,  
[\cf \eqs(\ref{HSC}) and (\ref {HSn3})]
\begin{eqnarray}
\hat{H}_{S_1}&\ug& \Delta \hat \alpha^\dag\hat \alpha\,,\,\,\hat{V}_{S\!S_1}\ug G(\hat\sigma_-\hat \alpha^\dag\piu{\rm H.c.})\,,\,\,\label{Hlossy1}\\
\hat{W}_{S_1 n}&\ug& g\,(\hat \alpha\hat a_n^\dag\piu{\rm H.c.}),\label{Hlossy2}
\end{eqnarray}
hence both the $S$-$S_1$ and $S_1$-$R_n$ interaction take place under the rotating wave approximation (RWA).

To illustrate the dynamics of the collision model defined this way, we consider the zero-temperature dynamics occurring when $S$ is initially in its excited state, while both $S_1$ and the ancillas are  in their vacuum states (hence, in particular, $\eta\ug|0\rangle\langle 0|$). The total number of excitations of $\mathcal S$-$\mathcal R$ is conserved at each collision, as follows from the form of \eqs(\ref{Hlossy1}) and (\ref{Hlossy2}). Given the considered initial state, the process thus takes place within the single-excitation sector of the Hilbert space.  Based on this, we use a compact notation according to which the overall initial state is denoted by $|10{\bf 0}\rangle$, where the first two quantum numbers refer to $S$ and $S_1$, respectively, while ${\bf 0}$ refers to the reservoir ancillas (indicating that they are all in the vacuum states). 
With the same notation, $|01{\bf 0}\rangle$ is the the state with the single excitation localised in  $S_1$ while
$|001_i\rangle$ is the state with the single excitation localised on the  $i$th reservoir ancilla $S_i$. At any step $n$, the joint state thus reads
\begin{equation}
  \ket{\Phi^{(n)}}=       \varepsilon^{(n)}\ket{10{\bf 0}}+
                             \beta^{(n)}  \ket{01{\bf 0}}+
\sum_{i=1}^n\lambda^{(n)}_{i}\ket{001_i}\,.\label{phin}
\end{equation}
Here, the superscript ``$n$" labels the $n$th time step, while the subscript ``$i$" on $\lambda$ labels the $i$th ancilla.
Note that the last sum in the equation above runs up to $i\ug n$ since at the end of the $n$th step ancillas labeled by index $i\!\ge\!n\piu1$ are still unexcited.

State $|\Phi^{(n)}\rangle$ is connected with $|\Phi^{(n-1)}\rangle$ as 
$\ket{\Phi^{(n)}}{=} e^{-i\hat{W}_{S_1 n}\tau} e^{-i(\hat{H}_{S_1}\!+\!\hat V_{S\!S_1})\tau}\ket{\Phi^{(n-1)}}$ [\cf\eqs(\ref{Usn2}), (\ref{Hlossy1}) and (\ref{Hlossy2})]. In Appendix A, we show that this allows to express coefficients $\{\varepsilon_n,\beta_n\}$ as linear functions of $\{\varepsilon_{n-1},\beta_{n-1}\}$ through the 2$\times$2 transformation matrix ${\bf M}$ given by

\begin{equation}
{\bf M}\ug  e^{-i\frac{\Delta}{2}  \tau}\left(
\begin{array}{cc}
z & -i\frac{G}{\Omega}   \sin (\Omega\tau)\\
-i \frac{G}{\Omega} \sin(\Omega\tau) \cos(g\tau)&  z^*\cos (g \tau )
\end{array} 
\right),\label{Mmatrix}
\end{equation}

where 
\begin{eqnarray}
\Omega&\ug&\frac{1}{2}\sqrt{\Delta^2\piu 4 G^2}\label{Omega}\,,\,\,\,\,\,z\ug  \cos(\Omega \tau )\piu i\frac{\Delta}{2\Omega} \sin (\Omega \tau )\,.
\end{eqnarray}
Upon iteration, 
\begin{eqnarray}
\left(\begin{array}{c}
 \varepsilon^{(n)} \\
\beta^{(n)}
\end{array}\right)= {\bf M}^n\!\left(\begin{array}{c}
 \varepsilon^{(0)} \\
\beta^{(0)}\label{epsnsol}
\end{array}\right)\,,
\end{eqnarray}
where in our case $\varepsilon^{(0)} \ug1$ while $\beta^{(0)} \ug0$.
\eq(\ref{epsnsol}) in particular allows to compute step-by-step the evolution of the excitation amplitude of $S$ up to any desired time $n\tau$. 

\begin{figure*}[ht!]
\begin{center}
\includegraphics[width=2\columnwidth]{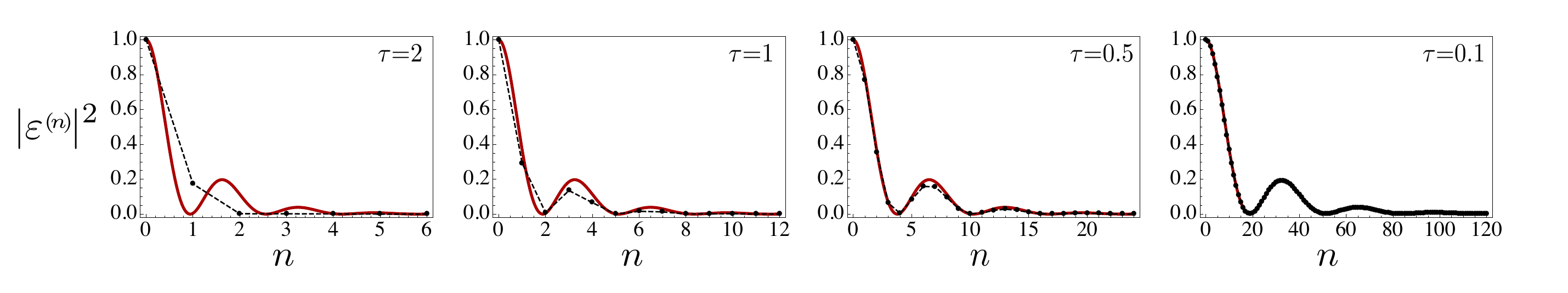}
\caption{Excited-state population $|\varepsilon^{(n)}|^2$ of $S$ against the step number $n$ in the case of the composite collision model specified by \eqs(\ref{Hlossy1}) and (\ref{Hlossy2}) for different values of the collision time $\tau$ (in units of $G^{-1}$) and for $g\ug\sqrt{G/\tau}$, $\Delta\ug0$. For each set value of $\tau$, the solid curve shows the behaviour of the excitation probability [\cf\eq(\ref{cossin})] predicted by master equation (\ref{KLME-JC}) for $t\ug n\tau$ and $\gamma\ug g^2\tau$.  
For $\tau\ug10^{-1}G^{-1}$ the solid curve is in fact indistinguishable from the exact discrete dynamics.} \label{fig-sim}
\end{center}
\end{figure*}

In \fig\ref{fig-sim}, we use \eq(\ref{epsnsol}) to illustrate how the discrete-step evolution of the $S$ excited-state population depends on the collision time $\tau$ in the paradigmatic case of zero detuning ($\Delta\ug 0$) and $g{=}\sqrt{G/\tau}$ (we use $G$ as the frequency unit and let $g$ be $\tau$-dependent in a way that $g^2\tau$ is fixed to $G$) . If $\tau$ is not short enough, a continuous-time approximation of the dynamics fails (see cases $\tau\ug 2 G^{-1}$ and $\tau\ug G^{-1}$ in \fig\ref{fig-sim}). Collision times of the order of $\tau\!\sim\!0.1G^{-1}$ or shorter are already enough to determine a smooth evolution as a function of the step number $n$. 
For the considered parameters, the $S$ dynamics in this limit exhibits damped oscillations.
These originate from the $S$-$S_1$ coupling Hamiltonian term $\hat V_{S\!S_1}$ [\cf\eqs(\ref{Hlossy1})], which in absence of reservoir would induce a continuous excitation exchange between $S$ and the auxiliary system $S_1$. The effect of the reservoir is to damp the amplitude of such energy exchange. These features can be explained by noting that for $\tau\!\ll\!G^{-1}$ the conditions required for master equation (\ref{MEC}) to hold (see Section \ref{MCM-multipartite}) are matched. Using that $\eta\ug|0\rangle\langle 0|$, the master equation  takes the explicit form [\cf\eqs(\ref{Hlossy1}) and (\ref{Hlossy2})]
\begin{eqnarray}
\dot\rho\ug\meno i\!\left[\Delta \hat \alpha^\dag\hat \alpha\piu G(\hat\sigma_-\hat \alpha^\dag\piu{\rm H.c.}),\rho\right]\piu\! \gamma\! \left(\!\hat{\alpha}\rho\hat{\alpha}^\dag\meno\frac{1}{2}[\hat {\alpha}^\dag\hat \alpha,\rho]_+\!\right)\!\,\,\,\,\,\,\,\,\,\,\,\,\label{KLME-JC}
\end{eqnarray}
with $\gamma\ug g^2\tau$ [in passing, note that condition \eqref{average} is fulfilled]. 
This is the well-known master equation (in the rotating frame) occurring in the damped Jaynes-Cummings (JC) model \cite{JC} describing the dynamics of a two-level atom of frequency $\omega_0$ coupled with rate $G$ to a single-mode  cavity of frequency $\omega_c$, where $\Delta\ug \omega_c\meno\omega_0$ is the detuning while $\gamma$ represents the cavity dissipation rate.

For $\rho(0)\ug |e\rangle_S\langle e||0\rangle_{S1}\langle 0|$,  the joint state of $\mathcal S$ and $\mathcal R$ at time $t$ must have the same form as \eq(\ref{phin}) with $\varepsilon^{(n)}\!\rightarrow\!\varepsilon(t)$, $\beta^{(n)}\!\rightarrow\!\beta(t)$ and $\lambda_i^{(n)}\!\rightarrow\!\lambda_i(t)$. This alongside master equation (\ref{KLME-JC}) then entail that $\varepsilon(t)$ obeys the integro-differential equation (see Appendix B)
\begin{eqnarray}
\dot{\varepsilon}=- \,G^2 \int_0^t {\rm d}t'\,e^{-i \left(\Delta-i\frac{\gamma}{2}\right)(t-t')}\varepsilon(t')\,,\label{eps-int}
\end{eqnarray}
the solution of which reading
\begin{eqnarray}
\varepsilon(t)\ug e^{-i\frac{\Delta}{2}t}e^{-\frac{\gamma}{4}t}  \left[\cos
   \left(\frac{\delta t}{2}\right)\!+\!i\,\frac{\omega_1 }{\delta}\sin
   \left(\frac{\delta t}{2}\right)\right]\label{cossin}\,
\end{eqnarray}
with
\begin{eqnarray}
\omega_1\ug \Delta-i\frac{\gamma}{2}\,,\,\,\,\,\delta\ug\sqrt{4
   G^2+\omega_1^2}\nonumber\,.
\end{eqnarray}
For $\Delta\ug0$ (zero detuning), $\delta\ug\sqrt{G^2\meno\gamma^2/4}$. Hence, for $\gamma\!\le\!2G$ and $\gamma\!>\!2G$ the excitation probability $|\varepsilon(t)|^2$ respectively exhibits damped oscillations and a monotonic (in general non-exponential) decay \cite{open}. In \fig \ref{fig-sim}, we compare the time evolution of the excitation probability of $S$ predicted by \eq(\ref{cossin}) corresponding to master equation (\ref{KLME-JC}) with the exact discrete dynamics of the collision model computed through \eq(\ref{epsnsol}). The agreement between these is excellent for collision times shorter than $\tau\!\sim\! 10^{-1}G^{-1}$.

\subsection{Connection with a microscopic environmental model}\label{connect}

Consider the microscopic environmental model defined by the Hamiltonian
\begin{eqnarray}
\hat H\!_{AF}\ug\omega_0\,\hat\sigma_+\hat \sigma_-\piu\!\sum_k \omega_k\,\hat a^\dag_k\hat a_k\piu\sum_k \mu_k \left(\hat\sigma_-\hat a^\dag_k \!+\!  \hat \sigma_+\hat a_k\right)\!,\,\,\,\,\,\,\,\,\label{HAF}
\end{eqnarray}
describing a two-level atom  $A$ of frequency $\omega_0$ in dissipative contact (under RWA) with a bath of bosonic modes (field), labelled by index $k$, with frequency $\omega_k$ and bosonic annihilation (creation) operator $\hat a_k$ ($\hat a_k^\dag$) and atom-mode coupling rate $\mu_k$. In the continuous limit, $\omega_k\!\rightarrow\!\omega$, $\mu_k\!\rightarrow\!\mu(\omega)$ and $\sum_k\!\rightarrow\!\int \!d\omega \,\rho(\omega)$ with $\rho(\omega)$ the field density of states. 

Consider the spontaneous emission process with the atom initially in its excited state and the bath modes initially in their vacuum state. Let $\varepsilon(t)$ be the probability amplitude to find the atom in its excited state at time $t$. Given Hamiltonian (\ref{HAF}), it can be shown \cite{open} that $\varepsilon(t)$ is governed by the general integro-differential equation
\begin{equation}
\dot{\varepsilon}=-\int_0^t{\rm d}s\,\left[\int d\omega J(\omega)e^{i(\omega_0 -\omega)(t-s)}\right]\varepsilon(s)\,,\label{epsdot}
\end{equation}
where $J(\omega)\ug[\mu(\omega)]^2\rho(\omega)$ is the spectral density.

Now, for a  Lorentzian spectral density given by
\begin{eqnarray}
J(\omega)\ug \frac{\Gamma_0 }{2\pi}\,\frac{\kappa^2}{(\omega\meno\omega_0\meno\Delta)^2+\kappa^2}\label{lorentzian}\,,
\end{eqnarray}
it can be shown that the {\it exact} solution for $\varepsilon(t)$ coincides with \eq(\ref{cossin}) provided that
\begin{eqnarray}
\gamma\ug2\kappa \,,\,\,\,\,\,\,G\ug\sqrt{\frac{\Gamma_0\kappa}{4}}\label{cond-JC}\,.
\end{eqnarray}
As long as the open dynamics of the two-level system is concerned, this in fact establishes an equivalence (first pointed out by Garraway \cite{garraway}) between the environmental model (\ref{HAF}) and the master equation (\ref{KLME-JC}) of the damped JC model. This is intuitively clear once in \eq(\ref{lorentzian}) $\omega_0\piu\Delta$ is interpreted as the resonance frequency of a cavity mode and $\kappa$ as the bandwidth of a lossy cavity. Within our framework, given the previously shown correspondence between master equation (\ref{KLME-JC}) and the collision model in \eqs(\ref{Hlossy1}) and (\ref{Hlossy2}), we can thus establish a correspondence between such composite collision model (in the continuous-time limit) and the microscopic environmental model in \eq(\ref{HAF}). Starting from the latter, we can thereby construct an associated composite (bipartite) collision model defined by \eqs(\ref{Hlossy1}), (\ref{Hlossy2}) and the parameters: $\Delta$,
\begin{eqnarray}
g\ug\sqrt{\frac{2\kappa}{\tau}}\,,\,\,\,G\ug\sqrt{\frac{\Gamma_0\kappa}{4}}\,,\label{mapping-lor}
\end{eqnarray}
where we used $\gamma\ug g^2\tau$ in combination with \eqs(\ref{cond-JC}).

To summarise, given the environmental microscopic model in \eq(\ref{HAF}), in the case of a Lorenztian spectral density [\cf\eq(\ref{lorentzian})], one can construct a composite collision model through \eqs(\ref{Hlossy1}), (\ref{Hlossy2}) and (\ref{mapping-lor}) which, in the continuous-time limit, 
reproduces the same open system dynamics.

\section{Random telegraph noise and pure dephasing}\label{dephasing}

In the next instance of composite bipartite collision model that we consider, $S$, $S_1$ and $R_n$ are all qubits. 
By definition [\cf \eqs(\ref{HSC}) and (\ref {HSn3})],
\begin{eqnarray}
\hat{V}_{S\!S_1}{=} G \;\hat K_S \hat K_{S_1}\,,\,\,\,\,\hat{W}_{S_1 n}&\ug& g(\hat \sigma_{1-}\hat \sigma_{n+}\piu\hat \sigma_{1+}\hat \sigma_{n-})\label{HRTN2}\,
\end{eqnarray}
with $\hat K_S$ ($\hat K_{S_1}$) a Hermitian operator on $S$ ($S_1$).
We take as initial state of each ancilla, a thermal state 
$\eta=\frac{1}{2}(1{-}\xi)\ket{0}\bra{0}{+}\frac{1}{2}(1{+}\xi)\ket{1}\bra{1}$ with 
$\xi\ug\tanh(\beta)$ and $\beta$ the $\mathcal R$'s inverse temperature. 

In the continuous-time limit (see Section \ref{MCM-multipartite}), the collision model defined this way gives rise to the master equation for the $S$-$S_1$ state
\begin{eqnarray}
\dot\rho{=}&&{-} i\left[
G\;\hat K_S \hat K_{S_1},\rho\right]\piu
\Gamma_+\!\left(\!\hat{\sigma}_{1-}\rho \,\hat{\sigma}_{1+}\meno\frac{1}{2}[\hat {\sigma}_{1+}\hat \sigma_{1-},\rho ]_+\!\right)\nonumber\\
&&+\,\Gamma_-\!\left(\!\hat{\sigma}_{1+}\rho \,\hat{\sigma}_{1-}\meno\frac{1}{2}[\hat {\sigma}_{1-}\hat \sigma_{1+},\rho ]_+\!\right)
\,,\,\,\,\,\,\,\,\,\,\,\,\label{KLME}
\end{eqnarray}
where $\Gamma_\pm\ug \gamma(1\pm\xi)/2$ with $\gamma\ug g^2\tau$.

We will consider next the collision models arising from two different choices of operators $\hat K_S$ and $\hat K_{S_1}$.

\subsection{Random telegraph noise}

In this first instance, we set $\hat K_S\ug \hat H_S/G$, $\hat K_{S_1}\ug\hat\sigma_{1z}$ and $\xi{=}0$ (hence the ancillary initial states are all  maximally mixed). 
Operator $\hat H_S$ can be interpreted as a Hamiltonian operator on $S$.
Let $\ket{\pm}_{S_1}$ be
the state of $S_1$ such that $\hat H_S\ket{\pm}_{S_1}\ug \pm\ket{\pm}_{S_1}$. 
Tracing over $S_1$, the $S$ reduced state is given by $\rho_S(t){=}\rho_{S+}(t){+}\rho_{S-}(t)$ with
$\rho_{S\pm}(t){=} _{S_1}\!\langle \pm|\rho(t)|\pm\rangle_{S_1}$.
In the continuous-time limit, the master equation (\ref{KLME}) gives rise to the following pair of coupled master equations
\begin{equation}
\dot{\rho}_{S\pm}{=}-i \left[\pm \hat H_S,\rho_{S\pm}\right]\pm\frac{1}{t_c}(\rho_{S-} \meno \rho_{S+})\,.\label{MERTN}
\end{equation}
which describe the well-known dynamics of a quantum system $S$ subjected to {\it random telegraph noise}  
\cite{vankampen} in the case of a single bistable fluctuator featuring a correlation time $t_c{=}2/g^2\tau$. In such a case, $\pm \hat H_S$ is the
Hamiltonian corresponding to the fluctuator's classical state labeled by ``$\pm$", in turn defining one of the two possible trajectories along which $S$ can evolve.

The possibility to derive a random telegraph noise qubit dynamics from a bipartite master equation of the form \eq(\ref{KLME}) was pointed out in \rref
\cite{Saira2007}.

\subsection{Pure Dephasing}\label{puredephasing}

Let us now set $\hat K_{S}{=}\hat\sigma_z$ and $\hat K_{S_1}{=}\hat\sigma_{1x}$ with states $\ket{0}$ ($\ket{1}$) 
being the $S$ state such that $\hat \sigma_z \ket{0}\ug \ket{0}$ ($\hat \sigma_z \ket{1}\ug- \ket{1}$) and  $\rho_0$ 
the initial state of $S$ . The populations $\langle m|\rho_0|m\rangle$ for $m\ug0,1$ 
will be clearly unaffected by the collision process due to the dispersive nature of the $S$-$S_1$ coupling while,
 as shown in Appendix \ref{Appdephasing}, the coherences, at the  $n$-th step are 
$$\langle 0|\rho_n|1\rangle\ug \langle 1|\rho_n|0\rangle^*\ug f_n \langle 0|\rho_0|1\rangle\,,$$
where the step-dependent dephasing factor $f_n$ reads
\begin{eqnarray}\label{fn}
f_n{=}&&\frac{1}{2}\left[\left(\frac{(c_g{+}1) c_G {+} \kappa} {2}\right)^{\!\!n}\!\!\!{+}\! \left(\frac {(c_g{+}1) c_G {-} \kappa} {2}\right)^{\!\!n}\right] \\
&&	{+}	      \frac {(1{-}c_g) c_G}{\kappa}\frac{1}{2}
\left[\! \left(\frac {(c_g{+}1) c_G {+} \kappa} {2}\right)^{\!\!n}\!\!\!{-}\! \left(\frac {(c_g{+}1) c_G {-} \kappa} {2}\right)^{\!\!n}\right]\,\,\,\,\,\,\,\,\,\nonumber
\end{eqnarray}

with
$$c_g=\cos(2 g \tau)\,,\,\,c_G=\cos(2G\tau)\,,\,\,\kappa {=}\sqrt{(c_g{-}1)^2 c_G^2-4 c_g s_G^2}\,\,.$$

To carry out the continuous-time limit, in line with Section \ref{MCM-internal}, we expand up to the second order in $g\tau$ and first order in $G\tau$. 
This way, neglecting terms proportional to $\sim \!G^2g^2$ and taking $n\tau\rightarrow t$ and $g^2\tau\rightarrow \gamma$, the continuous-time limit of $(1{-}c_g) c_G/\kappa$ turns out to be $g^2\tau/\kappa_c$ while $(c_g{+}1) c_G {\pm} \kappa$ becomes $2(1{-}g^2\tau\pm\kappa_c\tau)$, where 
\begin{equation}\label{kappac}
\kappa_c{=}\sqrt{\gamma ^2{-}4 G^2}\,.
\end{equation} 
{With these approximations the step-dependent decoherence factor \eqref{fn} takes the continuous-time form}
\begin{equation}
\!\!f(t){=}e^{-\gamma t} \left[\cosh (\kappa_c t){+}\frac{\gamma\sinh (\kappa_c t)}{\kappa_c}\right]\,.
\label{decohfactor}\end{equation} 
This result can also be derived through a direct solution of master equation \eqref{KLME}.

Similarly to Subsection \ref{connect}, also the present collision model can be associated with a corresponding microscopic environmental model yielding the same $S$ open dynamics.
To see this, consider a qubit $S$ dispersively coupled to a bosonic reservoir according to the Hamiltonian 
\begin{equation}
\hat H{=}\omega_0 \hat \sigma_{+}\hat\sigma_- {+} \sum_k\omega_k \hat a_k^\dagger \hat a_k{+}\sum_k \mu_k\hat \sigma_z(\hat a_k+\hat a_k^\dagger),
\end{equation}
where the difference with respect to \eq(\ref{HAF}) is that the interaction is now dispersive. As in Subsection \ref{connect}, the reservoir spectral density in the continuous limit is given by $J(\omega)\ug[\mu(\omega)]^2\rho(\omega)$.

This model can be solved exactly \cite{Luczka1990,Palma1996,open}, the corresponding master equation for the qubit (in the interaction picture) reading
\begin{equation}
\dot{\rho}_S=\gamma(t)\left(\sigma_z\rho_S\,\sigma_z-\rho_S\right)
\end{equation}
If the environment is initially in the vacuum state, the time-dependent dephasing rate $\gamma(t)$ takes the form
\begin{equation}
\gamma(t)=\int_0^\infty \!\!{\rm d}\omega \sin(\omega t)\frac{J(\omega) }{\omega}\,.
\label{gamma}
\end{equation}
Accordingly, the coherences decay as $\langle 0 |\rho(t)|1\rangle\ug e^{-\Lambda(t)}\langle 0 |\rho(0)|1\rangle)$ with $\Lambda(t)$ related to $\gamma(t)$ as $\Lambda(t){=}2 \int_0^t\!\gamma(t'){\rm d}t'$.
\eq(\ref{gamma}) shows that $J(\omega)/\omega$ is the Fourier-Sine transform of $\gamma(t)$. Correspondingly,
\begin{equation}
J(\omega)=\omega\int_0^\infty \!\!{\rm d}t \sin(\omega t) \gamma(t)\,.\label{Jomde}
\end{equation}
In the continuous-time limit of our collision model, we can identify $\Lambda(t)\ug-\log[f(t)]$ and thereby $\gamma(t)\ug1/2 \dot\Lambda(t)\ug-1/2\dot f(t)/f(t)$. Using next \eq(\ref{Jomde}) with the
help of \eq(\ref{decohfactor}), we thus find that for $\gamma{>}2G$ the equivalent spectral density of our collision model is given by
\begin{equation}
J(\omega){=}\sum _{j=0}^{\infty}\dfrac{(\gamma{-}\kappa_c )^j}{ (\gamma{+}\kappa_c )^{j{+}2}}\, \dfrac{4 G^2 \kappa_c  (j{+}1)^2}{(j{+}1)^2{+}\omega ^2/(4 \kappa_c^2)}
\;.
\label{Jdeph}
\end{equation}
We thus find that our collision model yields the same reduced dynamics of $S$  obtained 
from a microscopic environmental model where the reservoir spectral density consists of a series of Lorentzian-shaped distributions (with positive weights).
Note that all of these are centered at the same frequency with a width that increases with index $j$.
In the limit where the coupling rate $G$ is much smaller than the decay rate $\gamma$, only the first term of the sum dominates in a way that the spectral density reduces to a single Lorentzian. 

\section{Extension to the multipartite case} \label{MCM-multipartite-multi}

So far we have considered composite collision models where the system $\mathcal S$ comprises the very open system $S$ and a single auxiliary system $S_1$. In this section, we present an extension of the collision model to account for multiple auxiliary systems. Furthermore, in this new scenario, we will in addition allow each ancilla to be multipartite.
Such extension  enables a collision model-based description of certain open dynamics that cannot be captured in the simple bipartite case, as we will show in the next section.

\begin{figure}[h!]
\includegraphics[width=\columnwidth]{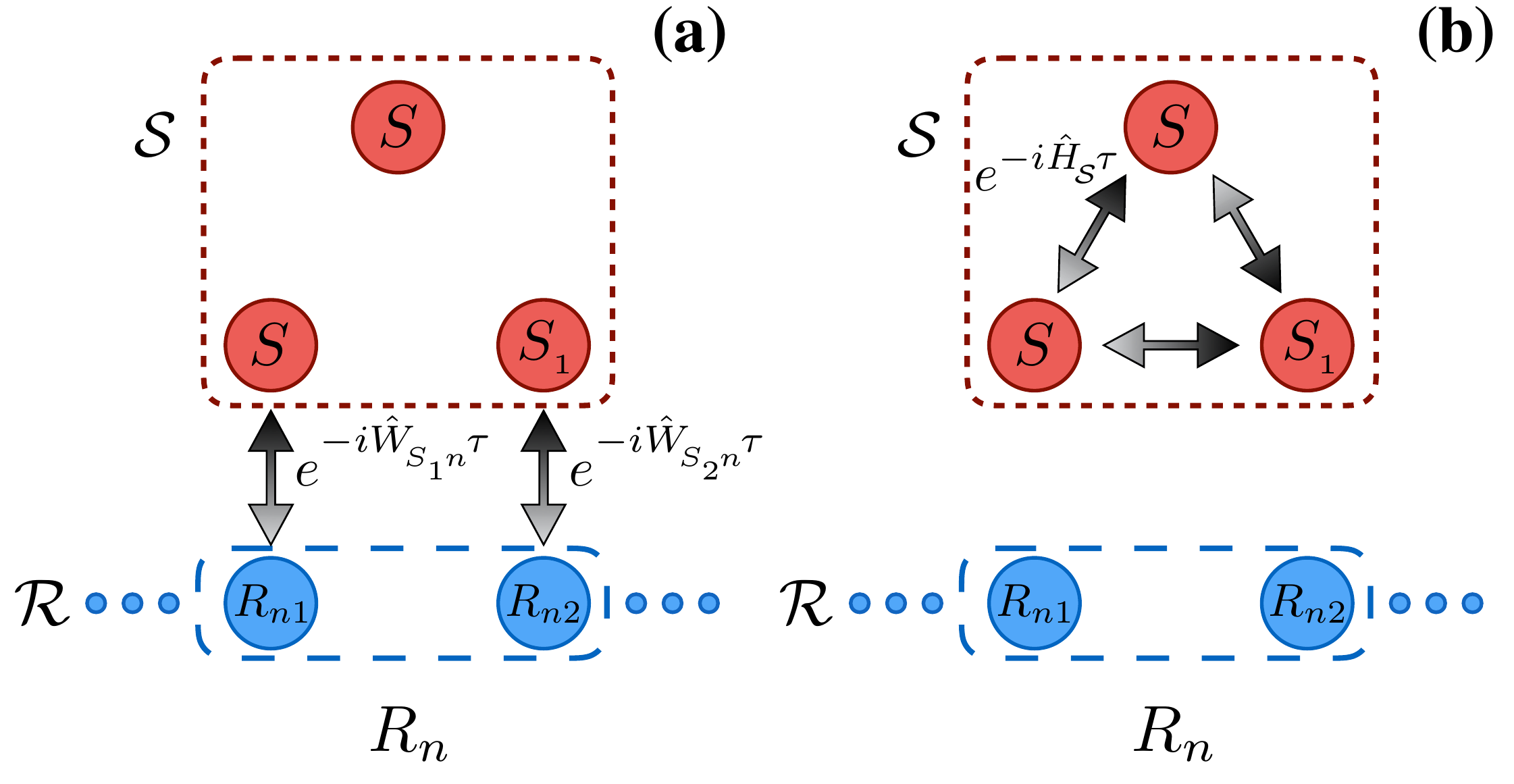}
\caption{\label{multifig}System-ancilla collision (a) and intra-system collision (b) in a composite tripartite collision model. System $\mathcal S$ is tripartite comprising the very open system $S$ under study and the auxiliary systems $\{S_{1}$, $S_2\}$, while ancilla $R_n$ is bipartite comprising sub-ancillas $\{R_{n1}$, $R_{n2}\}$. Note that a system-ancilla collision (a) occurs through pairwise interactions between the auxiliary systems and the corresponding reservoir sub-ancillas.}
\end{figure}

Both $\mathcal S$ and {\it each} reservoir ancilla are now assumed to be {\it multipartite}. Specifically, $\mathcal S$ comprises  $N\piu1$ subsystems $S,\,S_1,\,S_2,\,...,S_N$ with $S$ embodying the very open system under study and where $\{S_i\}$ are auxiliary systems. Furthermore, the $n$th reservoir ancilla $R_{n}$ is $N$-partite, its subsystems -- referred to as ``sub-ancillas" in the following -- being $\{R_{n1},R_{n2},...,R_{nN}\}$.  The free hamiltonian of $\mathcal S$ is now defined by [\cf\eq\eqref{HSC}]
\begin{equation}
\hat {H}_\mathcal S=\sum_{i=1}^{N}\,\left(\hat {H}_{S_i}+\hat V_{S\!S_i}\right)+\sum_{i< j}\hat V_{S_i\!S_j}\,,\label{HSC-multi}
\end{equation}
where $\hat {H}_{S_i}$ is the free Hamiltonian of subsystem $S_i$, $\hat V_{S\!S_i}$ is the interaction Hamiltonian of $S$ and $S_i$, while $\hat V_{S_i\!S_j}$ [not appearing in \eq(\ref{HSC})] describes the $S_i$-$S_j$ coupling between different auxiliary systems.
The $\mathcal S$-$R_n$ interaction [\cf\eq\eqref{HSn3}] is generalized as
\begin{equation}
\hat {H}_{\mathcal{S}n }=\sum_{i=1}^{N}\,\hat {W}_{{S_i}{n_i}}\ug g_i\hat {w}_{{S_i}{n_i}} \label{HSn3-multi}
\end{equation}
with $\hat {W}_{{S_i}{n_i}}$ the interaction Hamiltonian of subsystem $S_i$ and {\it sub}ancilla $R_{n_i}$. Note that $\hat V$ operators describe interactions internal to the system $\mathcal S$, while the $\hat W$'s correspond to system-ancilla interactions. Also, note that the latter ones take place only between subsystems and sub-ancillas labeled by corresponding indexes. The bipartite composite model of Section \ref{MCM-multipartite} is retrieved in the special case $N=1$. A sketch of a composite tripartite collision model, corresponding to $N\ug2$, is given in \fig\ref{multifig}.

We assume the initial state of each ancilla  to be the product state $\eta\ug\otimes_{i=1}^N\xi_i$, with $\xi_i$ the initial state of each sub-ancilla, and that [\cf\eqs(\ref{average}) and (\ref{HSn3-multi})]
  \begin{equation}
{\rm Tr}_{R_{n_i}}\!\{ \hat W_{{S_i}{n_i}}\xi_i\}\ug0\,.\label{average2-multi}
 \end{equation}
This entails ${\rm Tr}_{R_n} \!\{\hat H_{\mathcal{S}n }\eta\}\ug0$. Under the above conditions, an identity analogous to \eq(\ref{drho2}) holds.

Once $\hat{H}_{\mathcal {S}n}$ is expressed through \eq(\ref{HSn3-multi}), we note that the resulting cross terms in the $\hat W$ operators vanish because of \eq\eqref{average2-multi}. In the continuous-time limit, we thus end up with a master equation for $\rho$ (state of $\mathcal S$) that reads
\begin{eqnarray}
\frac{d\rho}{dt}=-i\,[\hat{\mathcal H}_{\mathcal S}, \rho]\mathcal+ \sum_{i=1}^N\mathcal L_{S_i}(\rho)
\end{eqnarray}
with
\begin{eqnarray}
\mathcal L_{S_i}(\rho)&\!=\!&\gamma_i\!\sum_{\mu\nu} \left\{\!\!\hat{A}_{\mu\nu}^{(i)}\,\rho\left(\hat{A}_{\mu\nu}^{(i)}\right)^\dag\!\!\meno\frac{1}{2}\!\left[\left(\hat{A}_{\mu\nu}^{(i)}\right)^\dag\!\!\hat{A}_{\mu\nu}^{(i)},\rho\right]\!_{\!+}\!\!\right\},\nonumber \\
\hat{A}_{\mu\nu}^{(i)}&\!=\!&\sqrt{p_\nu}\langle \mu|\hat w_{{S_i}{n_i}}|\nu\rangle\,\nonumber\\
\gamma_i&\!=\!&{g_i^2\tau}\,,
\end{eqnarray}
where the initial sub-ancilla state is $\xi_i\ug\sum_m p_m |m\rangle_{\! i}\!\langle m|$ 
(here $|\mu\rangle$ and $|\nu\rangle$ are two generic elements of the orthonormal basis $\{|m\rangle\}$ in the Hilbert space of sub-ancilla $R_{n_i}$). The jump operator $\hat{A}_{\mu\nu}^{(i)}$ and its ${\rm H.c.}$ act in the Hilbert space of the auxiliary system $S_i$ with $i\ug1,...,N$.

In the following section, we show how the composite model defined above provides a collision-model description of a quantum emitter subject to a {\it multi-Lorentzian} spectral density.

\section{Multi-Lorentzian spectral density}\label{multi}

Here, we address a composite collision model which can be regarded as an extension of the model described in Section \ref{lossy} to the case $N\ug2$ (hence $\mathcal S$ is now tripartite). The model features {two} auxiliary systems $S_1$ and $S_2$, each modelled as a bosonic mode with associated annihilation (creation) operator $\hat \alpha_i^\dag$ ($\hat \alpha_i$) for $i\ug1,2$.  Correspondingly, each reservoir ancilla $R_n$ is now bipartite, consisting of sub-ancillas $R_{n1}$ and $R_{n2}$, each modelled as a bosonic mode of annihilation (creation) operator $\hat a_{ni}$ ($\hat a_{ni}^\dag$). The dynamics of this collision model is generated by the following Hamiltonian
[\cf \eqs(\ref{HSC-multi}) and (\ref {HSn3-multi})]
\begin{eqnarray}
\hat{H}_{S_i}&\ug& \Delta_i \hat \alpha_i^\dag\hat \alpha_i,\,\hat{V}_{S\!S_i}\ug G_i(\hat\sigma_-\hat \alpha_i^\dag\piu{\rm H.c.})\,,\label{Hmulti1}\\
\,\hat{V}_{S_1\!S_2}&\ug& c\, (\hat\alpha_1\hat \alpha_2^\dag\piu{\rm H.c.})\,,\,\,\,\,
\,\hat{W}_{{S_i}{n}}\ug g_i(\hat \alpha_i \hat a_{ni}^\dag\piu{\rm H.c.})\,\,\,\,\,\label{Hmulti2}
\end{eqnarray}
with $i\ug1,2$.
Note that, in general, the auxiliary systems $S_1$ and $S_2$ can be subjected to a mutual interaction (with associated coupling rate $c$).

We again restrict our analysis to the single excitation subspace, with  $S$   initially in its excited state, while $S_1$, $S_2$ and all the ancillas are initially in their ground state (hence now $\eta\ug|0\rangle_{R_{n1}}\!\langle 0|\!\otimes\!|0\rangle_{R_{n2}}\!\langle 0|$). Introducing a compact notation analogous to that employed in Section \ref{lossy}, the single-excitation initial state is thus denoted by $|100{\bf 0}\rangle$, the first three quantum numbers now referring to $S$, $S_1$ and $S_2$, respectively. The total number of excitations of $\mathcal S$-$\mathcal R$ is again conserved. Similarly to \eq(\ref{phin}), the joint state at an arbitrary step $n$ is now of the form 
\begin{eqnarray}
  \ket{\Phi^{(n)}}&=&       \varepsilon^{(n)}\ket{100{\bf 0}}+
                             \beta_1^{(n)}  \ket{010{\bf 0}}+
                             \beta_2^{(n)}  \ket{001{\bf 0}}\nonumber\\
                             &&+\sum_{i=1}^n\sum_{j=1}^2\lambda^{(n)}_{ij}\ket{001_{ij}}\,\nonumber
\end{eqnarray}
with $\beta_j^{(n)}$ $\left(\lambda_{ij}^{(n)}\right)$ the excitation probability amplitude of $S_j$ ($R_{nj}$) at the $n$th step.

By carrying out an analysis similar to the one carried on in section \ref{lossy}, one can show that in the continuos-time limit the present collision model 
yields for $S$ an open dynamics equivalent to that of a microscopic environmental model of the form (\ref{HAF}) where the
spectral density is a sum of two Lorentzian functions.

In line with \rref\cite{garraway}, we consider the two cases respectively specified by [see \eqs(\ref{Hmulti1}) and (\ref{Hmulti2})] (a) $c{=}0$ and (b) $G_2{=}0$, $\Delta_{1,2}{=}\Delta$.
In the case (a), we obtain that the equivalent spectral density is the sum of two spectral densities with positive weights

$$J(\omega){=}\sum_{i=1,2}\dfrac{4G_{i}^2/\gamma_{i}}{2\pi}\dfrac{(\gamma_{i}/2)^2}{(\omega -2\Delta_{i})^2+(\gamma_{i}/2)^2}$$ 
with $\gamma_i{=}g_i^2\tau$. Hence, 
the rate $\gamma_i/2$ and ratio $4G^2_i/\gamma_i$ give respectively the width and maximum of
each distribution.   
In the case (b), instead, under the condition ${\gamma_1{-}\gamma_2}{>}2c$, the equivalent spectral density turns out to be
\begin{equation}
J(\omega){=}\left(\frac{\kappa_{+}}{2\pi}\dfrac{\lambda_{+}^2}{(\omega -2\Delta)^2+\lambda_{+}^2}\right){-}\left(\frac{\kappa_{-}}{2\pi}\frac{\lambda_{-}^2 }{(\omega -2\Delta)^2+\lambda_{-}^2}\right)\nonumber\,,
\end{equation}
where
\begin{eqnarray}
\lambda_\pm{=}&&(\gamma_1{+}\gamma_2\pm\chi)/4 \nonumber\\
\kappa_\pm{=}&&\frac{2 G^2 \left(8 c^2 \left(\chi{\mp} 2 \gamma_2\right){\pm} \left(\gamma_1{-}\gamma_2\right) \gamma_2 \left(\gamma_1{-}\left(\gamma_2{\pm} \chi \right)\right)\right)}
{\chi ^2 \left(4 c^2+\gamma_1 \gamma_2\right)}\nonumber
\end{eqnarray}
with $\chi{=}\sqrt{(\gamma_1{-}\gamma_2)^2{-}16c^2}$. 

Such a combination of Lorentzian distributions (with weights of opposite signs) can be used as a simplified model of a reservoir featuring a band-gapped spectrum \cite{open,garraway}.

\section{Conclusions}\label{Conclusions}

In this paper, we have introduced a class of quantum collision models, which we called composite collision models. Their definition is inspired by multipartite Lindblad-type master equations to describe non-Markovian dynamics, where the open system under study is coherently coupled to one or more auxiliary systems which are in turn in contact with Markovian baths. 

In one such collision model, the very open system under study $S$ (undergoing non-Markovian dynamics in general) is coupled to one or more auxiliary (in general mutually interacting) systems $\{S_i\}$, which in turn interact with the reservoir ancillas. We have presented a comprehensive discussion of the continuous-time limit in which the collision model is effectively described by a master equation, in particular the conditions on the collision time and Hamiltonian parameters to fulfill.

We have shown that this collision model-based framework can accommodate some known relevant instances of non-Markovian dynamics, such as an atom decaying in a lossy cavity, a qubit subjected to random telegraph or purely dephasing noise and a quantum emitter in dissipative contact with a reservoir featuring a spectral density that is the sum of two Lorentzian distributions. 

It was also illustrated that some specific microscopic environmental models can interestingly be associated with suitably-built corresponding composite collision models yielding the same open system dynamics in the continuous-time limit. 

The theory presented here strengthens the role that collision models can play as an alternative, advantageous approach for tackling quantum non-Markovian dynamics. 

An open question left is whether collision models can be constructed to describe certain classes of non-Markovian dynamics that cannot be captured by the framework developed here, such as the decay of an atom in a photonic-band-gap medium where the corresponding reservoir spectral density exhibits van Hove singularities, or -- if not -- whether such impossibility can be given an insightful physical meaning \cite{PBG}.

\section*{Acknowledgements}
We acknowledge support from the EU Project QuPRoCs (Grant Agreement 641277). 
We  thank R.~McCloskey, M.~Paternostro and K.~Luoma for fruitful discussions.

\appendix

\section{}
To derive \eq(\ref{epsnsol}), which in particular yields the open dynamics of $S$ in the collision model of Section \ref{lossy}, we first define the unitaries $\hat U_{S_1 n}\ug e^{-i\hat{W}_{S_1 n}\tau}$ and $\hat U_{\mathcal S}\ug e^{-i(\hat{H}_{S_1}\!+\!\hat V_{S\!S_1})\tau}$. Next, we use that $\hat U_{S_1 n}$ acts on the $n$th ancilla and subsystem $S_1$ only, while $\hat U_{\mathcal S}$ acts only on $S$ and $S_1$. Also, either of these unitary operators does not change the total number of excitations. Thereby, in the single-excitation  sector of the total Hilbert space, the only three states to be affected by the application of $\hat{U}_{S_1 n }\hat{U}_{\mathcal S}$ are $\{|10{\bf 0}\rangle,\,|01{\bf 0}\rangle,\,|00\,1_{n}\rangle\}$, all the remaining ones being invariant. Thus, in virtue of \eq(\ref{phin}) for $n\!\rightarrow\!n-1$, 
\begin{eqnarray}
  \ket{\Phi^{(n)}\!}\ug  \hat{U}_{S_1\!R_n }\hat{U}_{\mathcal S}\!\!\left[  \!  \varepsilon^{(n-1)}\ket{10{\bf 0}}\piu
                             \beta^{(n-1)}  \ket{01{\bf 0}}\piu\!\sum_{i=1}^{n-1}\gamma^{(n-1)}_{i}\ket{001_i}\!\right]\!.\nonumber\\\label{phin2}
\end{eqnarray}
Based on \eqs(\ref{Hlossy1}) and (\ref{Hlossy2}), the effective matrix representation of $\hat{U}_{S_1 n }\hat{U}_{\mathcal S}$ in the subspace $\{|10{\bf 0}\rangle,\,|01{\bf 0}\rangle,\,|00\,1_{n}\rangle\}$ can be calculated as
\begin{widetext}
$${\bf U}_{S_1 n }\!{\bf U}_{\mathcal S}\ug\left(
\begin{array}{ccc}
 e^{-i\frac{\Delta}{2}  \tau} \left[\cos(\Omega \tau )\piu i\frac{\Delta}{2\Omega} \sin
  (\Omega \tau )\right] & \frac{G}{2\Omega}  e^{- i 
   \left(\frac{\Delta}{2}+\Omega\right)\tau } \left(1\meno e^{2i\Omega \tau  }\right)& 0 \\
 \frac{G}{4\Omega}  \left(1\piu e^{2 i g \tau }\right) \left(1\meno e^{2i \Omega \tau }\right)\!e^{- i 
   \left(\Omega\piu\tfrac{\Delta}{2}\piu g\right)\tau }& e^{-i\frac{\Delta}{2} 
   \tau } \cos (g \tau ) \left[ \cos(\Omega \tau )\meno i \tfrac{\Delta} {2\Omega}  \sin
  (\Omega \tau )\right]& -i \sin (g \tau ) \\
 \frac{G}{4\Omega} \left(1\meno e^{2 i g \tau }\right) \left(1\meno e^{2i  \Omega \tau }\right) \!e^{- i 
   \left(\Omega\piu\tfrac{\Delta}{2}\piu g\right)\tau } & \,\,-ie^{-i\frac{\Delta}{2} \tau }
   \sin (g \tau ) \left[  \cos(\Omega \tau )-i\frac{\Delta}{2\Omega} \sin(\Omega \tau ) 
 \right] & \cos (g \tau )
\end{array} 
\right),$$
\end{widetext}
where $\Omega\ug\sqrt{\Delta^2\piu 4 G^2}/2$ [see \eqs(\ref{Omega})]. This alongside \eq(\ref{phin}) thus yield
\begin{widetext}
\begin{eqnarray}
\varepsilon^{(n)}&\ug& e^{-i\frac{\Delta}{2}  \tau} \left[\cos(\Omega \tau )\piu i\frac{\Delta}{2\Omega} \sin
  (\Omega \tau )\right]\varepsilon^{(n-1)}\piu \frac{G}{2\Omega}  e^{- i 
   \left(\frac{\Delta}{2}+\Omega\right)\tau } \left(1\meno e^{2i\Omega \tau  }\right)\beta^{(n-1)}\,,\label{epsn}\\
   \beta^{(n)}&\ug&  \frac{G}{4\Omega}  \left(1\piu e^{2 i g \tau }\right) \left(1\meno e^{2i \Omega \tau }\right)\!e^{- i 
   \left(\Omega\piu\tfrac{\Delta}{2}\piu g\right)\tau }\varepsilon^{(n-1)}\piu e^{-i\frac{\Delta}{2} 
   \tau } \cos (g \tau ) \left[ \cos(\Omega \tau )\meno i \tfrac{\Delta} {2\Omega}  \sin
  (\Omega \tau )\right]\beta^{(n-1)}\,,\label{betan}\\
     \lambda_n^{(n)}&\ug&  \frac{G}{4\Omega} \left(1\meno e^{2 i g \tau }\right) \left(1\meno e^{2i  \Omega \tau }\right) \!e^{- i 
   \left(\Omega\piu\tfrac{\Delta}{2}\piu g\right)\tau }  \varepsilon^{(n-1)} \meno ie^{-i\frac{\Delta}{2} \tau }
   \sin (g \tau ) \left[  \cos(\Omega \tau )-i\frac{\Delta}{2\Omega} \sin(\Omega \tau ) 
 \right]  \beta^{(n-1)}.
\end{eqnarray}
\end{widetext}
\eqs(\ref{epsn}) and (\ref{betan}) can be expressed in matrix form as
$$\left(\begin{array}{c}
 \varepsilon^{(n)} \\
\beta^{(n)}
\end{array}\right)\ug {\bf M}\left(\begin{array}{c}
 \varepsilon^{(n-1)} \\
\beta^{(n-1)}
\end{array}\right)\,,$$
where the 2$\times$2 matrix ${\bf M}$ [\cf\eq(\ref{Mmatrix})] is the upper-left 2$\times$2 block of matrix ${\bf U}_{S_1\!R_n }\!\!{\bf U}_{\mathcal S}$.

\section{Derivation of \eq(\ref{cossin})}

For $\eta\ug|0\rangle\langle0|$ and $\rho(0)\ug |e\rangle_S\langle e||0\rangle_{S1}\langle 0|$, due to the conservation of the total number of excitations [\cf\eqs(\ref{Hlossy1}) and (\ref{Hlossy2})] the joint $\mathcal S$-$\mathcal R$ state at time $t$ has the form
\begin{equation}
  \ket{\Phi(t)}=       \varepsilon(t)\ket{10{\bf 0}}+
                             \beta(t)  \ket{01{\bf 0}}+
\sum_{i=1}^n\lambda_{i}(t)\ket{001_i}\,.
\end{equation}
Upon trace over $\mathcal R$ of $ \ket{\Phi(t)}\bra{\Phi(t)}$, the $\mathcal S$'s density matrix then reads
\begin{eqnarray}
  \rho(t)&=&      | \varepsilon(t)|^2\ket{10}\!\bra{10}\piu
                             | \beta(t)|^2\ket{01}\!\bra{01}\nonumber\\
                 &&            +(\varepsilon(t)[\beta(t)]^*\ket{10}\bra{01}\piu{\rm H.c.})\nonumber\\
                      &&       +(1\meno|\varepsilon(t)|^2\meno |\beta(t)|^2)\ket{00}\!\bra{00} \,.             
\end{eqnarray}
Plugging this into master equation (\ref{KLME-JC}) yields the following set of equations
\begin{eqnarray}
\frac{{\rm d} }{{\rm d}t}|\varepsilon|^2&\ug& i G (\varepsilon\beta^* \meno  \varepsilon^*\beta)\nonumber\,,\\
\frac{{\rm d} }{{\rm d}t}|\beta|^2&\ug&-i G(\varepsilon \beta^*  \meno \varepsilon^* \beta ) \meno \gamma |\beta|^2\nonumber\,,\\
\frac{{\rm d} }{{\rm d}t}\left(\varepsilon\beta^*\right)&\ug&-\frac{\gamma}{2}\varepsilon\beta^*+i \left[G (|\varepsilon|^2 \meno|\beta|^2 )\piu \Delta
    \varepsilon\beta^* \right]\,.
\end{eqnarray}
\\
It is easily checked that these are equivalent to the system of differential equations in the excitation amplitudes
\begin{eqnarray}
\dot{\varepsilon}&\ug& -iG\beta\,,\label{eq1}\\
\dot{\beta}&\ug&  -i\left(\Delta-i \frac{\gamma}{2}\right)\beta-i G\varepsilon\label{eq2}\,.
\end{eqnarray}
Solving \eq(\ref{eq2}) as a function of $\varepsilon(t)$ under the initial condition $\beta(0)\ug0$ yields
\begin{eqnarray}
\beta(t)=-i\, G\! \int_0^t {\rm d}t'\,e^{-i \left(\Delta-i \frac{\gamma}{2}\right)(t-t')}\varepsilon(t')\,,
\end{eqnarray}
which when replaced in \eq(\ref{eq1}) gives rise to the integro-differential equation (\ref{eps-int}).
The solution (\ref{cossin}) of \eq(\ref{eps-int}) can be worked out by taking the Laplace transform  of each equation side so as to end up with an algebraic equation in the Laplace transform of $\varepsilon(t)$. Once the inverse Laplace transform is computed, \eq(\ref{cossin}) is obtained.

\section{}\label{Appdephasing}

At step $n$, the $S$-$S_1$ joint state (i.e., the $\mathcal S$'s one) reads
\begin{eqnarray}\label{deph_n-map}
\rho_{n}{=}{\rm Tr}_{R_n}\!\!\left\{\hat{U}_{\mathcal S n}\rho_{n-1}\,\eta\,\hat {U}_{\mathcal S n}^\dag\right\}\ug\mathcal{F}[\rho_{n-1}]\ug\mathcal F^n[\rho_{0}]\;\;\;\;\;\;
\end{eqnarray}
with $\hat{U}_{\mathcal Sn}\ug \exp(-i\hat{V}_{S\!S_1}\tau)\exp(-i\hat{W}_{S_1\!R_{n}}\tau)$ [\cf\eq(\ref{HRTN2}) and Subsection \ref{puredephasing}], where we have defined
the bipartite quantum map $\mathcal F$ acting on $S$ and $S_1$.

Next, we take as local operator basis for $S$ and $S_1$ the set of Hermitian operators
\begin{eqnarray}
\!\!\!\!\hat{\mathcal G}_{\alpha0}{=}\mathbb{I}_\alpha/\sqrt{2},\,\hat{\mathcal G}_{\alpha1}{=}\sigma_{\alpha x}/\sqrt{2},\,
\hat{\mathcal G}_{\alpha2}{=}\sigma_{\alpha y}/\sqrt{2},\,\hat{\mathcal G}_{\alpha3}{=}\sigma_{\alpha z}/\sqrt{2}\nonumber\\
\nonumber
\end{eqnarray}
with $\alpha\ug S,S_1$. Accordingly, the bipartite operator basis for the joint $S$-$S_1$ system is given by $\hat{G}_{kj}{=}\hat{\mathcal G}_{Sk}{\otimes}\hat{\mathcal G}_{S_1j}$.

In this representation, map $\mathcal{F}$ corresponds to a $16{\times}16$ matrix $\bf F$, whose entries are given by $\mathbf{F}_{(kj,k'j')}{=}\text{Tr}\lbrace\hat{G}_{kj}\mathcal{F}[\hat{G}_{k'j'}]\rbrace$, while
the bipartite state $\rho_n$ is turned into the 16-dimensional column vector ${\bf r}_n$ defined by $r_{n,kj}\ug {\rm Tr}\{\hat G_{kj}\rho_n\}$
Matrix $\bf F$ and vector ${\bf r}_0$ are then computed as
\begin{widetext}
\begin{eqnarray}
\mathbf{F}&{=}&\left(
\begin{smallmatrix}
  1 & 0 & 0 & 0 & 0 & 0 & 0 & 0 & 0 & 0 & 0 & 0 & 0 & 0 & 0 & 0 \\
 0 & c_g & 0 & 0 & 0 & 0 & 0 & 0 & 0 & 0 & 0 & 0 & 0 & 0 & 0 & 0 \\
 0 & 0 & c_G c_g & 0 & 0 & 0 & 0 & 0 & 0 & 0 & 0 & 0 & 0 & 0 & 0 & -c_g s_G \\
 -s_g^2 & 0 & 0 & c_G c_g^2 & 0 & 0 & 0 & 0 & 0 & 0 & 0 & 0 & 0 & 0 & c_g^2 s_G & 0 \\
 0 & 0 & 0 & 0 & c_G & 0 & 0 & 0 & 0 & -s_G & 0 & 0 & 0 & 0 & 0 & 0 \\
 0 & 0 & 0 & 0 & 0 & c_G c_g & 0 & 0 & -c_g s_G & 0 & 0 & 0 & 0 & 0 & 0 & 0 \\
 0 & 0 & 0 & 0 & 0 & 0 & c_g & 0 & 0 & 0 & 0 & 0 & 0 & 0 & 0 & 0 \\
 0 & 0 & 0 & 0 & -c_G s_g^2 & 0 & 0 & c_g^2 & 0 & s_G s_g^2 & 0 & 0 & 0 & 0 & 0 & 0 \\
 0 & 0 & 0 & 0 & 0 & s_G & 0 & 0 & c_G & 0 & 0 & 0 & 0 & 0 & 0 & 0 \\
 0 & 0 & 0 & 0 & c_g s_G & 0 & 0 & 0 & 0 & c_G c_g & 0 & 0 & 0 & 0 & 0 & 0 \\
 0 & 0 & 0 & 0 & 0 & 0 & 0 & 0 & 0 & 0 & c_g & 0 & 0 & 0 & 0 & 0 \\
 0 & 0 & 0 & 0 & 0 & -s_G s_g^2 & 0 & 0 & -c_G s_g^2 & 0 & 0 & c_g^2 & 0 & 0 & 0 & 0 \\
 0 & 0 & 0 & 0 & 0 & 0 & 0 & 0 & 0 & 0 & 0 & 0 & 1 & 0 & 0 & 0 \\
 0 & 0 & 0 & 0 & 0 & 0 & 0 & 0 & 0 & 0 & 0 & 0 & 0 & c_g & 0 & 0 \\
 0 & 0 & 0 & -c_g s_G & 0 & 0 & 0 & 0 & 0 & 0 & 0 & 0 & 0 & 0 & c_G c_g & 0 \\
 0 & 0 & c_g^2 s_G & 0 & 0 & 0 & 0 & 0 & 0 & 0 & 0 & 0 & -s_g^2 & 0 & 0 & c_G c_g^2 \\
\end{smallmatrix}
\right)\,,\\
{\bf r}_0&{=}&\frac{1}{2}\left\{1,0,0, {\meno }1, \rho_0 ^{(01)}{+}\rho_0 ^{(10)},0,0, \meno (\rho_0 ^{(01)}{+}\rho_0 ^{(10)}), i (\rho_0 ^{(10)}{\meno }\rho_0 ^{(01)}),0,0,\meno  i(\rho_0 ^{(10)}{\meno }\rho_0 ^{(01)}), \rho_0 ^{(11)}{\meno }\rho_0 ^{(00)},0,0, \rho_0 ^{(00)}\meno \rho_0 ^{(11)}\right\}\nonumber
\end{eqnarray}
\end{widetext}
with $s_X{=}\sin(2 X \tau)$ and $c_X{=}\cos(2 X \tau)$ for $X{=}g,G$ and $\rho_0^{(ij)}\ug \langle i|\rho|j\rangle$ for $i,j\ug0,1$. 

Evaluating next matrix $\mathbf{F}^n$ and applying it on ${\bf r}_0$, one can calculate ${\bf r}_n$ and eventually return to the density-matrix description through $\rho_n{=}\sum_{kj}  r_{kj}\hat{G}_{kj}$. This way, we end up with \eq(\ref{fn}) where $f_n{=}\frac{1}{2}[(\mathbf{F}^n)_{(21,21)}{+}(\mathbf{F}^n)_{(31,31)}]$.

\end{document}